\documentclass[12pt,preprint]{aastex}

\newcommand\plottwovert[2]{\centering \leavevmode
 \includegraphics[width={0.6\columnwidth}]{#1} \hfil
 \includegraphics[width={0.6\columnwidth}]{#2}}

\begin{document}

%
%--put definitions here
%
\def\msun{${\rm M_{\odot}} \;$}
\def\be{\begin{equation}}
\def\ee{\end{equation}}
\def\gcc{gcm$^{-3}$}
\def\bi{\begin{itemize}}
\def\ei{\end{itemize}}
\def\Mo{$M_{\odot}$}
\def\Lo{$L_{\odot}$}
\def\s{s$^{-1}$}

\title{Simulations of magnetic fields in filaments}

\author{M.Br\"{u}ggen\altaffilmark{1}, M.Ruszkowski\altaffilmark{2,3},
A.Simionescu\altaffilmark{1}, M.Hoeft\altaffilmark{1}, C.Dalla Vecchia\altaffilmark{4,5}}

\altaffiltext{1}{International University Bremen, Campus Ring 1, 28759 Bremen, Germany}
\altaffiltext{2}{JILA, Campus Box 440, University of Colorado at Boulder, CO 80309-0440}
\altaffiltext{3}{{\it Chandra} fellow}
\altaffiltext{4}{Department of Physics, University of Durham, South Road, Durham DH1 3LE, England}
\altaffiltext{5}{Sterrewacht Leiden, PO Box 9513, 2300 RA Leiden, The Netherlands}

\begin{abstract}

The intergalactic magnetic field within filaments should be less
polluted by magnetised outflows from active galaxies than magnetic
fields in clusters. Therefore, filaments may be a better laboratory to
study magnetic field amplification by structure formation than galaxy
clusters which typically host many more active galaxies. We present
highly resolved cosmological AMR simulations of magnetic fields in the
cosmos and make predictions about the evolution and structure of
magnetic fields in filaments. Comparing our results to observational
evidence for magnetic fields in filaments suggests that amplification
of seed fields by gravitational collapse is not sufficient to produce
IGM fields. Finally, implications for cosmic ray transport are
discussed.

\end{abstract}


\section{Introduction}

Clusters of galaxies are known to host magnetic fields with strengths
that are of the order of $\mu$G and coherence scales that are of the
order of 10 kpc (\citealt{carilli:02,govoni04}). In cool cores of
clusters remarkably high fields have been found:
E.g. \cite{blanton:03} have found magnetic fields as high as 11 $\mu$G
in the cool core of A2052. Knowledge about cluster magnetic fields
comes from synchrotron and inverse Compton radiation as well as
Faraday rotation measurements. Based on analyses of rotation measure
maps in three cluster, \citet{vogt03} have derived spectral indices of
1.6 to 2.0 (\citealt{vogt05, ensslin03}). However, they probe mostly
smaller scales than those considered here and, to our knowledge, there
are no reliable observations on larger scales.

Meanwhile, the origin of cluster magnetic fields in the IGM remains
unclear. It has been suggested that they are primordial
(\citealt{murgia:04, banerjee03,dolag02, gnedin:00}), i.e. that a seed
field that has formed prior to recombination is subsequently amplified
by compression and turbulence. In \cite{dolag02}, the average fields
in the core of clusters at redshift 0 were found to range between
0.4-2.5 $\mu G$ if the initial seed fields at redshift 15 were of the
order of a nanogauss. Alternatively, it has been proposed that the
magnetic field was of protogalactic origin (\citealt{kulsrud97}) or
that it had been produced by a cluster dynamo
(\citealt{ruzmaikin89,roettiger:99}). Others conclude that magnetic
fields can be produced efficiently in shocks by the Weibel instability
(\citealt{medvedev:04}) or by small-scale plasma instabilities
(\cite{schekochihin:05}). Finally, it has been suggested that the IGM
has been magnetised by bubbles of radio plasma that are ejected by AGN
(\citealt{furlanetto:01}). A growing number of large, magnetised
bubbles is being discovered in clusters of galaxies
(\citealt{birzan04}) and simulations show that they may not be very
long-lived (\citealt{bruggen:05}). Presumably, the magnetised bubble
plasma gets mixed with the IGM and is thus likely to contribute to
present cluster fields. However, this form of AGN activity has only
been discovered in the center of massive clusters, so that the IGM in
filaments ought to be much less affected by magnetised plasma ejected
from AGN.\\

\cite{bagchi:02} have discovered diffuse radio emission from a large
network of filaments of galaxies that span several Mpc. Assuming that
this emission is synchrotron emission the minimum-energy argument
yields a minimum field strength of $B\sim 0.3\,\mu$G.

The purpose of this paper is to simulate the magnetic field in
filaments with sufficient resolution in order to make predictions
about their magnitude and structure. There have been few cosmological
simulations that include magnetic fields, and most of them focus
solely on galaxy clusters. So far there are no computations that can
make reliable predictions on field amplifications in filaments.

Obviously, this project poses significant computational challenges as
we have to cover a wide range of scales. In order to simulate
structure formation on a cosmologically relevant scale, the box has to
be many Mpc across. At the same time we wish to resolve the thin
filaments well enough to make statements about the magnetic field
evolution within them. In an attempt to bridge this large range of
scales we used the adaptive-mesh PPM code FLASH that provides full
support for particles and cosmology.

\section{Simulation}

The initial conditions for our simulations are the publicly available
conditions of the Santa Barbara cluster (\citealt{frenk:99}) at
redshift $z=50$. Consequently, our cosmological parameters are those
of the Santa Barbara cluster, i.e. $H_0=50$ km \s Mpc$^{-1}$,
$\Omega_m=1$, $\Omega_{\Lambda}=0$.  The cluster perturbation
corresponds to a 3$\sigma$ peak in density smoothed over 10 Mpc. We
assumed standard cosmic composition, i.e. the Hydrogen and Helium
fractions were 0.7589 and 0.2360, respectively. Star formation has
been neglected.\\

Our computational domain was a cubic box of side $L=64 h^{-1}$ Mpc.
   FLASH is a modular block-structured AMR code, parallelised using
   the Message Passing Interface (MPI) library. FLASH solves the
   Riemann problem on a Cartesian grid using the Piecewise-Parabolic
   Method (PPM) and, in addition, includes particles that represent
   the dark matter. Our simulation included 2097152 dark matter
   particles with a mass of $7.8 \cdot 10^9 M_{\odot}$ each. We chose
   a block size of $16^3$ zones and used periodic boundary
   conditions. The minimal level of refinement was set to 4 which
   means that the minimal grid contains $16 \cdot 2^{(4-1)} = 128$
   zones in each direction. The maximum level of refinement was 10,
   which corresponds to an effective grid size of $16 \cdot 2^{(10-1)}
   = 8192$ zones or an effective resolution of $7.8 h^{-1}$ kpc. This
   was the maximum that we could afford computationally.\\

The mesh is refined and derefined automatically depending on the dark
matter density. It was set up such that no more than 8 dark matter
particles occupied one computational cell. We first ran a simulation
with only 8 levels of refinement, then identified three regions that
contained filaments, and reran the simulation with 10 levels of
refinement. In addition to refining on the dark matter we enforced
full refinements in the regions that contain the filaments. However,
the dark matter distribution, represented by the particles, is not
refined.

%Fig.~\ref{fig1} shows the block boundaries in a number of slices
%displaying the gas density and gives an impression of the range of
%scales covered in this simulation.

In order to achieve satisfactory accuracy even on small scales we
decided to implement a passive magnetic field solver in FLASH that
keeps the divergence of the magnetic field very small. The price for
this is that, currently, we are unable to compute the full MHD
equations. However, we find that the magnetic fields in the ICM will
rarely be dynamically important and, at this level, they are described
with sufficient accuracy by a passive field solver. \\

We chose to implement the passive field solver described by
\cite{Pen03}, which solves the magnetic field on a staggered mesh. It
is straightforward to show that if the magnetic field is evolved on a
staggered instead of a centred grid, the magnetic field will remain
divergence-free provided that it was divergence-free to start with
(Evans \& Hawley 1988). Every time step the magnetic field is evolved
using a total-variation diminishing (TVD) scheme that solves

\begin{equation}
\partial {\mathbf B} = \nabla \times ({\mathbf v \times B}) - \frac{3}{2}\frac{\dot{a}}{a}{\mathbf B} \ ,
\end{equation}
where ${\mathbf B}= {\mathbf B}_{\rm physical}/a$ and $a$ is the scale-factor.
The thus evolved field is then interpolated with a third-order scheme
to the cell centres.\\

Initially we set up a divergence-free field using the vector
potential. The magnetic vector potential $\tilde{\mathbf A}(k)$ was
composed of 512 Fourier modes with random phases and amplitudes that
were drawn from a Gaussian distribution. $\tilde{\mathbf A}$ was then
scaled as $k^{-\alpha}$, where $k$ is the wavenumber of the mode. This
leads to a magnetic power spectrum of $P_B(k)\, dk\sim
k^{2(2-\alpha)}\,dk$. Here, $\alpha$ was chosen to be 2. The vector
potential was then Fourier-transformed into physical space with a Fast
Fourier Transform. The field was normalised such that the physical
mean field at redshift 50 was $3\cdot 10^{-11}$ G.\\

In agreement with what \cite{dolag02} have found using SPH
simulations, we verified that the resulting magnetic fields are very
insensitive to the topology of the initial field at $z=50$. Even if
one starts with an initially uniform magnetic field, the results are
not significantly different. The field at any time obviously depends
on the strength of the initial field, but since the field is only
evolved passively, the normalisation is arbitrary.

\section{Results and Discussion}

\begin{figure*}[htp]
\includegraphics[width={6in}]{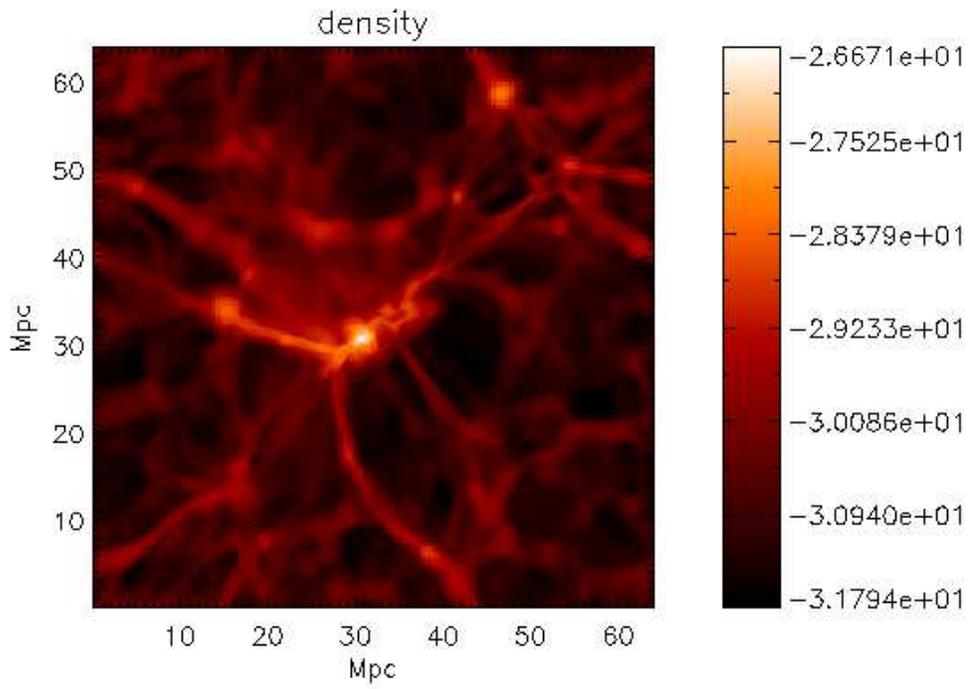}
\caption{Logarithmic density plots in slices through the center of the
computational box at $z=0.5$. Density units are cgs.}
\label{fig1}
\end{figure*}

\begin{figure*}[htp]
\includegraphics[width={6in}]{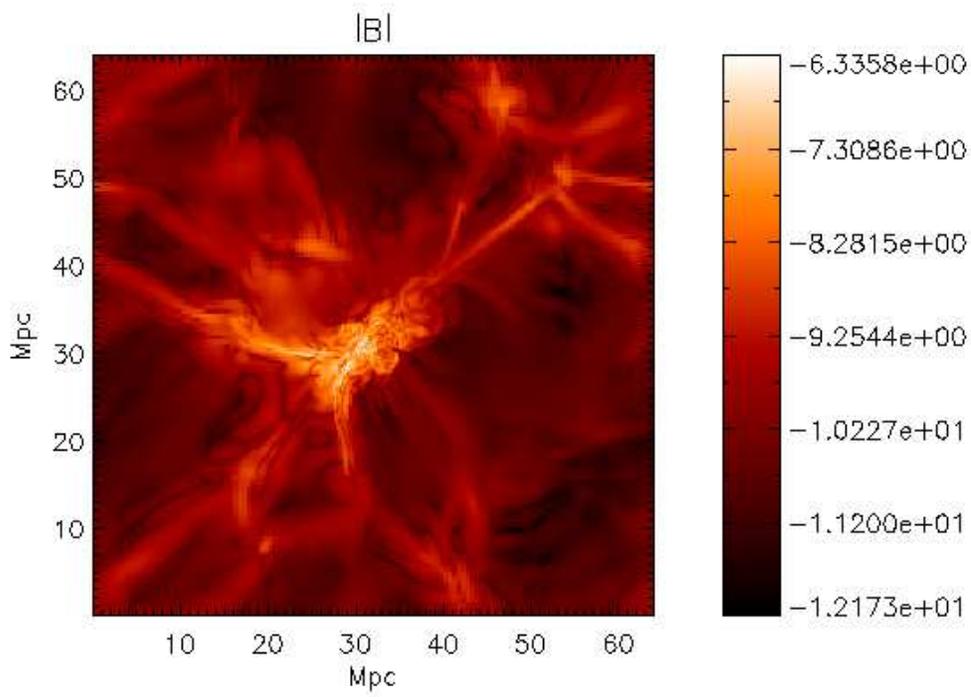}
\caption{Logarithmic plots of the magnitude of the magnetic field in slices through the center of the
computational box at $z=0.5$.}
\label{fig2}
\end{figure*}

In Fig.~\ref{fig1} we show the density distribution in a slice through
the center of the computational box where a massive cluster has
formed.  ($r_{200}= 2.6$ Mpc, total mass $\sim 1.1\cdot 10^{15}
M_{\odot}$) The cosmic web with its filaments is clearly
visible. Fig.~\ref{fig2} displays the magnitude of the magnetic field
in the same same slice as in Fig.~\ref{fig1}. Evidently the magnetic
field correlates with density as one would expect from the
conservation of magnetic flux which implies $B\propto
\rho^{2/3}$. However, we find that in some regions the magnetic field
is amplified significantly beyond $B\propto \rho^{2/3}$ which is what
one would expect from compression alone. This is caused by shear flows
and occurs mainly in the outskirts of the cluster and near accretion
shocks.

The mass-averaged field in the center (within 0.1 $r_{\rm vir}$) of
the cluster increases by a factor of $\sim 7000$. The correlation
length is of the order of 30 kpc and the modulus of the rotation
measure goes up to 1000 rad m$^{-2}$. This agrees with findings by
\cite{dolag02} who use SPH simulations of clusters, which is a
fundamentally different computational technique from the one used
here. \\

\begin{figure*}[htp]
\includegraphics[width={6in}]{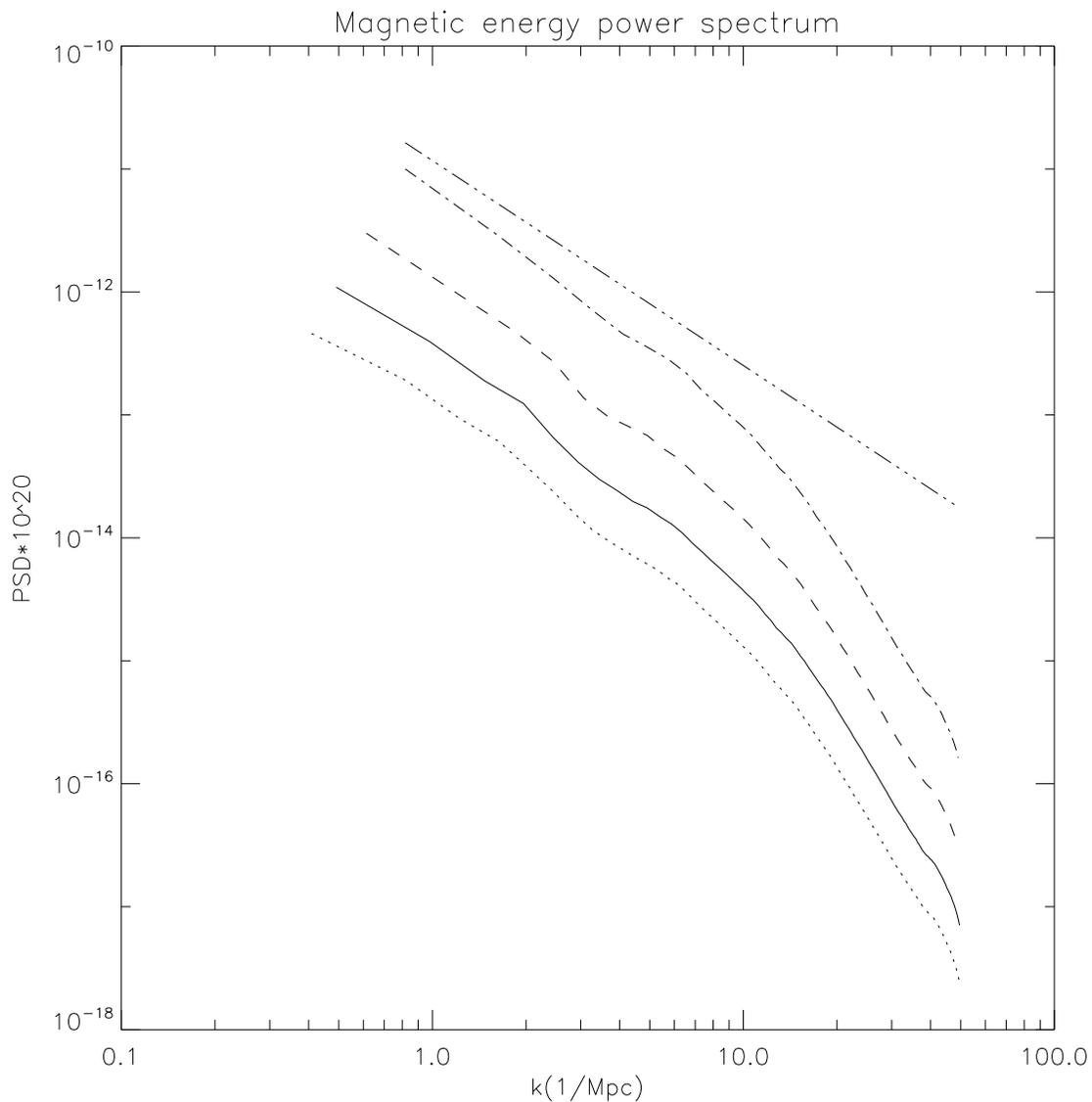}
\caption{Three-dimensional power spectrum of the magnetic field energy in the galaxy
cluster that forms at the center of the cluster. The dashed line represents a power law of index -5/3 for comparison. The lines correspond to different box sizes for which the power spectrum was calculated (dotted line: 15.36 Mpc, solid line: 12.80 Mpc, dashed line: 10.24 Mpc, dot-dashed line: 7.68 Mpc).}
\label{fig3}
\end{figure*}

In Fig.~\ref{fig3} we show the three-dimensional power spectrum of the
magnetic field energy $\epsilon\sim B^2$ for various box sizes centred
on the cluster. Until numerical effects near the resolution limit set
in, the power spectrum is well described by a power-law with an index
of $\sim 5/3$. Thus, the mere amplification of seed fields by
gravitational collapse and shearing motions reproduces cluster fields
with a power spectrum that is in very good agreement with observations
(\citealt{vogt05}). Similar slopes were found in \cite{rordorf:04}.

\begin{figure*}[htp]
\includegraphics[width={6in}]{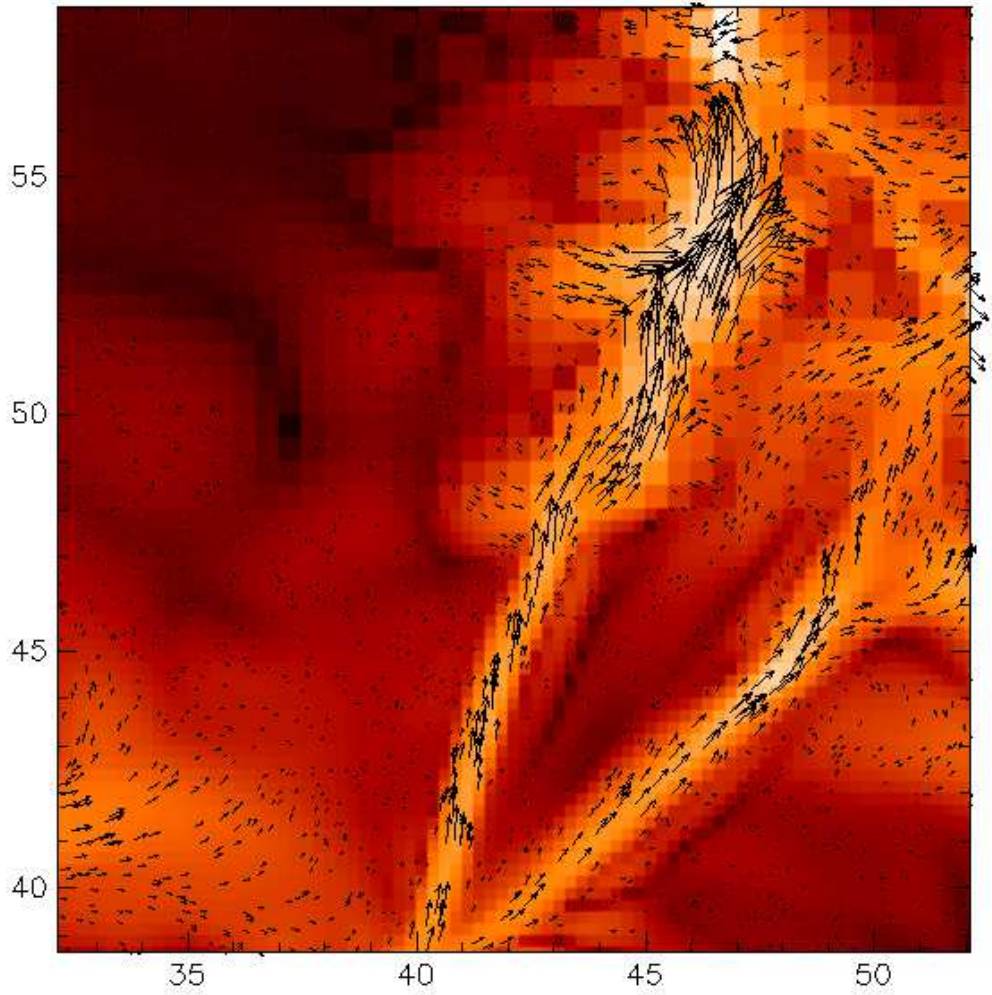}
\caption{Magnetic field in a slice through two filaments in our
computational box. The colors represent the logarithm of the magnitude of
the magnetic field and the tickmarks denote Mpc. The lines show the orientation of the magnetic
field in the plane of the plot.}
\label{fig4}
\end{figure*}

In the filaments the magnetic field increases by a factor of $\sim
10-300$ between a redshift of 50 and 0.5.  In
Fig.~\ref{fig4} we zoom into a region that contains two typical
filaments. The mass-averaged temperature in the filaments is $\sim
10^6$ K. Shown are the magnitude and orientation of the magnetic
field. In the filaments the magnetic field goes up to $\sim 3$ nG
which corresponds to an amplification by a factor of 100. One should
note that our grid simulations are maximally refined in the filaments
and are thus expected to yield more reliable results than SPH
simulations.\\

If indeed filaments host magnetic fields of the order of $B\ge
0.3\,\mu$G, as suggested by \cite{bagchi:02}, this would require seed
fields of the order of $10^{-9}$ G at $z\sim 50$. Seed fields of this
order of magnitude may be hard to produce (see also
\cite{dolag:05}). Consequently, this would call for alternative
origins for IGM magnetic fields (\citealt{kronberg:04}). Future
observing campaigns with instruments such as LOFAR and the SKA are
expected to shed some light on magnetic fields in the early
universe.\\

As is evident from Fig.~\ref{fig4}, we find that magnetic fields are
predominantly oriented in a direction parallel to the filaments. This
is true for all filaments in our simulation volume. As material
accretes onto the filaments, the magnetic fields are compressed
parallel to the filaments and, as clusters and groups form, the
filaments are stretched. This field topology could have several
interesting implications: For one, the diffusion length of cosmic rays
along filaments may be much greater than out of the filaments.  This
makes filaments efficient traps for cosmic rays and could help to
explain the diffuse radio emission seen along the entire length of the
filament in \cite{bagchi:02}. Secondly, heat conduction is likely to
be more efficient along filaments than elsewhere. This could lead to
heat being conducted from the outskirts of the clusters into the
filaments.\\

Moreover, we have calculated the cumulative volume filling fraction of
the magnetic field. It shows that only ~1\% of the volume is filled
with magnetic fields above $1\:{\rm nG}$. Our filling curve is close
to the results of \cite{dolag:05}. As a consequence UHECR deflection
of more than $1^\circ$ can only happen if a cosmic ray passes a
cluster region. We found that for 99\% of trajectories chosen randomly
through our box protons with $E\ge 10^{20}$ eV would be deflected by
less than $0.2^\circ$. These results are well consistent with first
positive indications of UHECR clustering
(\citealt{farrar:05}). Upcoming UHECR detectors as the Pierre Auger
experiment may probe the large-scale magnetic fields directly.\\

\begin{figure*}[htp]
\plottwovert{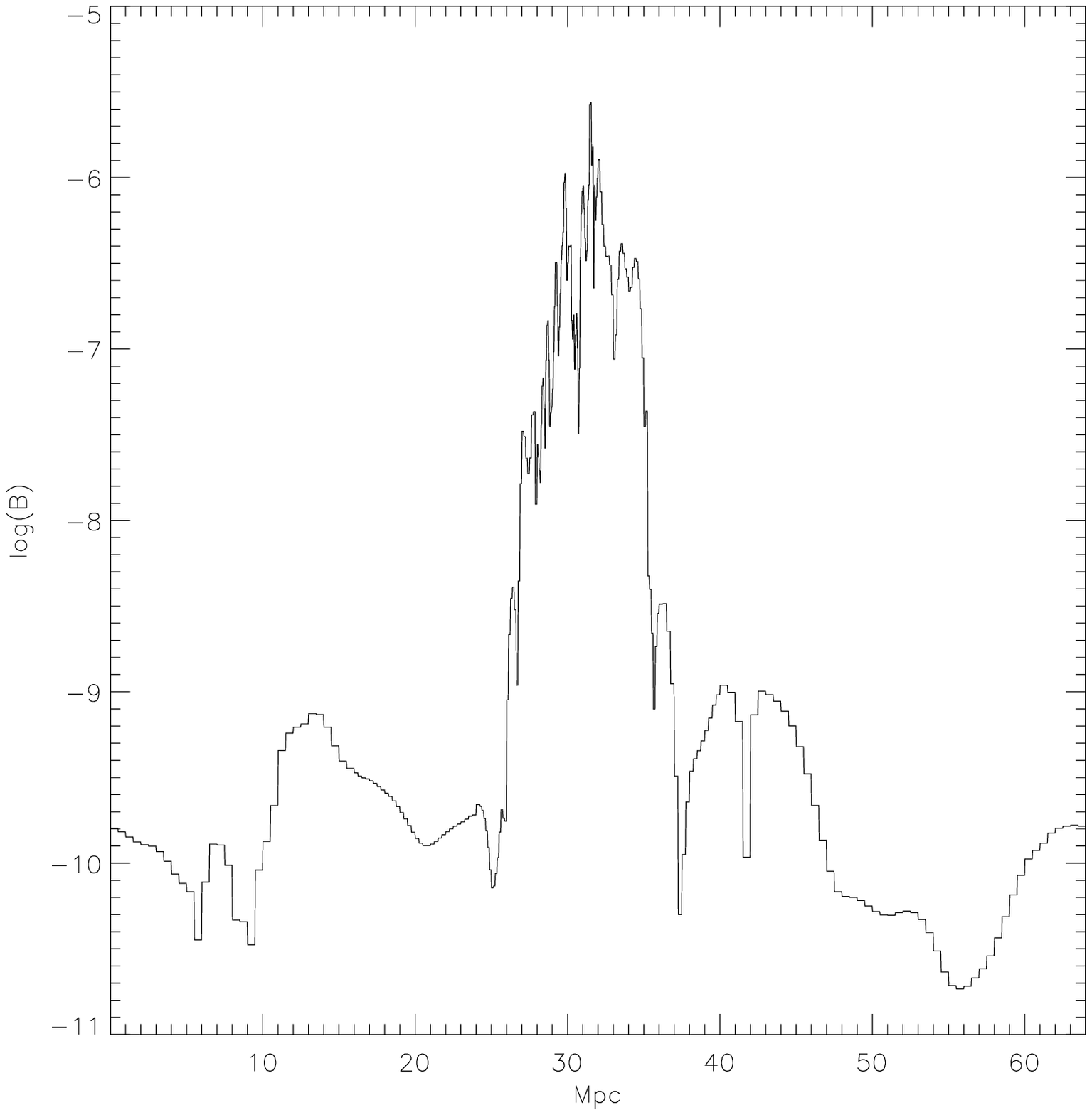}{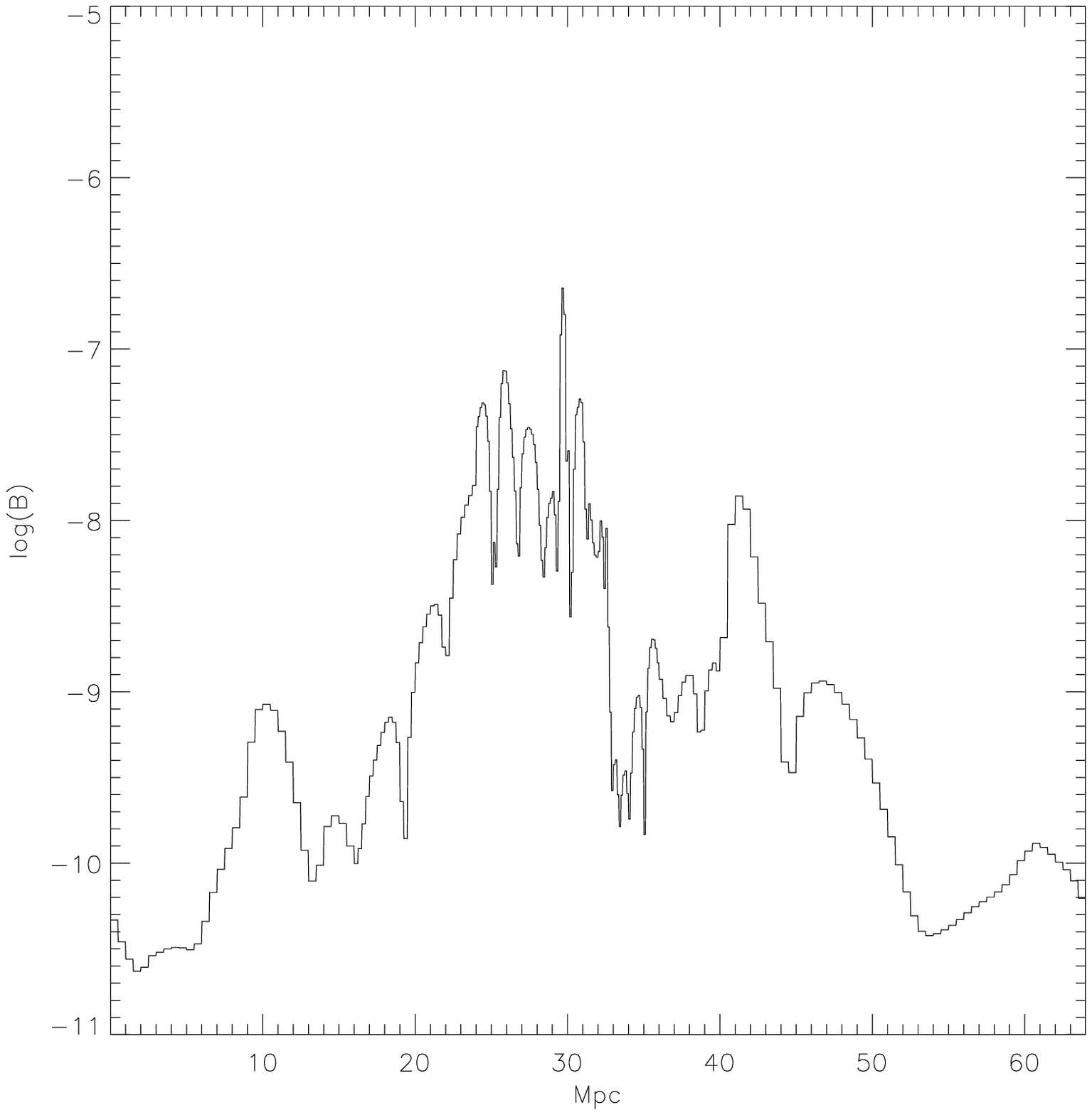}
\caption{Profile of the magnetic field along a line that goes through
the center of the cluster (top) and along a line that cuts through a
filament (bottom).}
\label{fig5}
\end{figure*}

Finally, in Fig.~\ref{fig5} we plot the magnetic field strength along
a line that goes through the center of the cluster and on a line that
cuts through the most massive filament in our computational box. The
relative strengths of the fields in clusters and filaments are
evident.\\

Obviously, in these kinds of simulations the question of grid
convergence arises. We have run our simulation at three different
levels of refinement and compared the results. While the field
amplification typically increases with the resolution, we found only a
very small difference between runs with 9 and 10 levels. However, one
should note that the resolution of the underlying dark matter density
and velocity field is restricted by the dark matter particle mass,
which is not refined. \\

Summary: We have produced a cosmological AMR simulation with a passive
magnetic field solver on the basis of the FLASH code. In agreement
with work by others, we find that within the cluster seed fields are
amplified by a factor of $\sim 7000$ between $z=50$ and today. We find
that cluster magnetic fields follow a power spectrum with index
$5/3$. Within filaments the amplification is typically $10-300$ and
the fields are aligned with the filament. The magnetic field of the
IGM correlates roughly linearly with temperature.

\vspace{0.5cm}
\noindent {\sc acknowledgement }
 
MB gratefully acknowledges support by DFG grant BR 2026/2 and the
supercomputing grant NIC 1658. This material is based upon work
supported by the National Science Foundation under the following NSF
programs: Partnerships for Advanced Computational Infrastructure,
Distributed Terascale Facility (DTF) and Terascale Extensions:
Enhancements to the Terascale Facility. The software used in this work
was in part developed by the DOE-supported ASCI/Alliance Center for
Astrophysical Thermonuclear Flashes at the University of Chicago. MR
acknowledges the support from NSF grant AST-0307502 and NASA through
{\it Chandra} Fellowship award number PF3-40029 issued by the Chandra
X-ray Observatory Center, which is operated by the Smithsonian
Astrophysical Observatory for and on behalf of NASA under contract
NAS8-39073.  

\bibliography{radio} 

\bibliographystyle{apj}

\end{document}